\documentclass[12pt,a4paper,final]{iopart}

\usepackage{iopams}
\usepackage{graphicx}
\usepackage[breaklinks=true,colorlinks=true,linkcolor=blue,urlcolor=blue,citecolor=blue]{hyperref}

\usepackage{sidecap}

\begin{document}

\title[Celebrating Haldane]{Celebrating Haldane's `Luttinger liquid theory'}

\author{Jean-S{\'e}bastien Caux}
\address{Institute for Theoretical Physics Amsterdam and Delta Institute for Theoretical Physics, University of Amsterdam, Science Park 904, 1098 XH Amsterdam, The Netherlands}
\ead{J.S.Caux@uva.nl}

\author{Cristiane Morais Smith}
\address{Institute for Theoretical Physics, Center for Extreme Matter and Emergent Phenomena, Utrecht University, Leuvenlaan 4, 3584 CE Utrecht, The Netherlands}
\ead{C.deMoraisSmith@uu.nl}

\begin{abstract}
This short piece celebrates Haldane's seminal J. Phys. C {\bf 14}, 2585 (1981) paper laying the foundations of the modern theory of Luttinger liquids in one-dimensional systems.
\end{abstract}

\section{Introduction}

In a pioneering paper published in J. Phys. C in 1981 \cite{1981_Haldane_JPC_14}, Duncan Haldane taught us that one-dimensional fermionic systems behave in a fundamentally different manner than those in the universality class described by Landau's Fermi liquid theory \cite{1957_Landau_JETP_3}-\nocite{1957_Landau_JETP_5}\cite{AbrikosovBOOK}.

\begin{center}
\begin{figure}
\includegraphics[width=8cm]{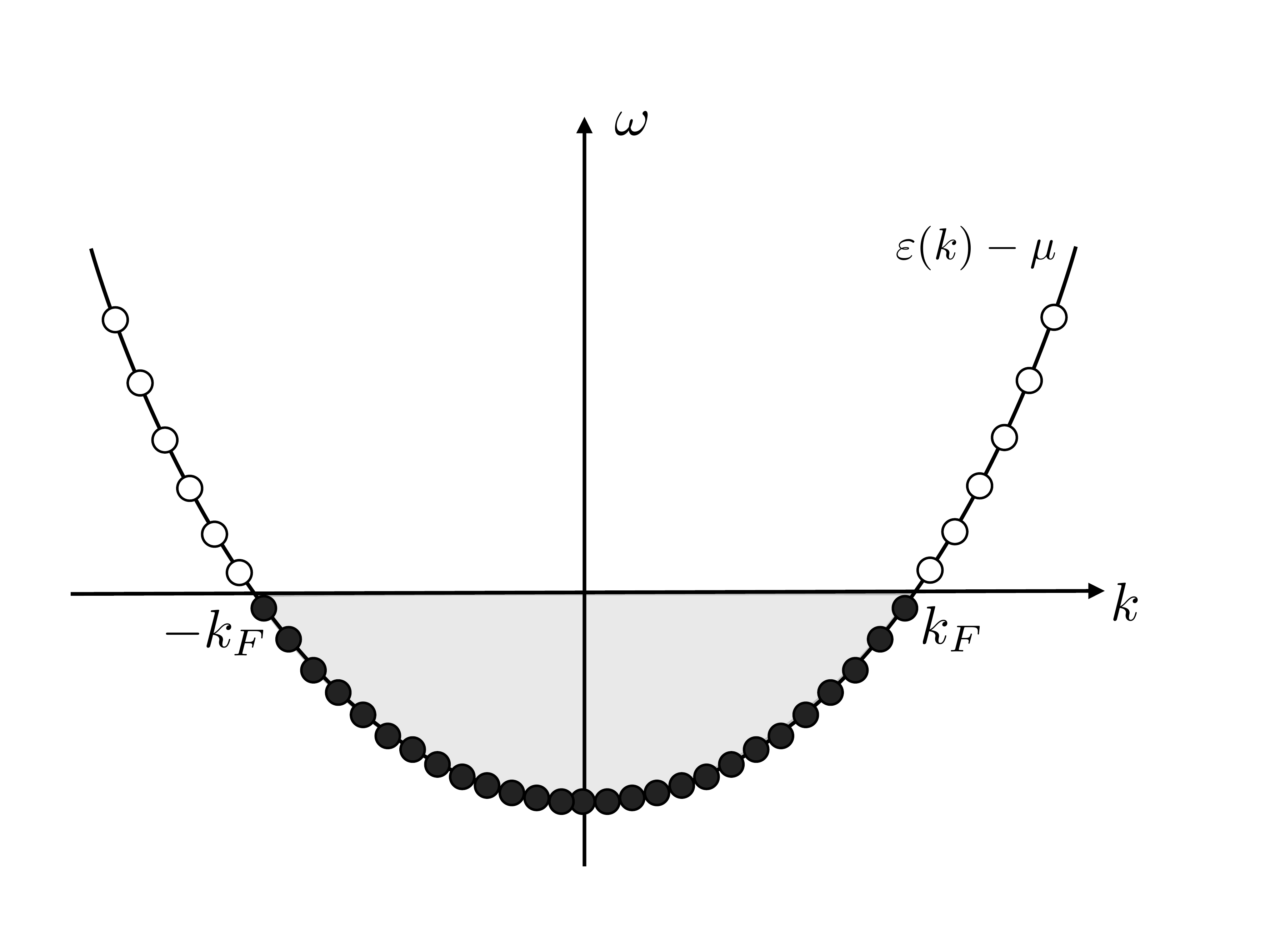}
\includegraphics[width=8cm]{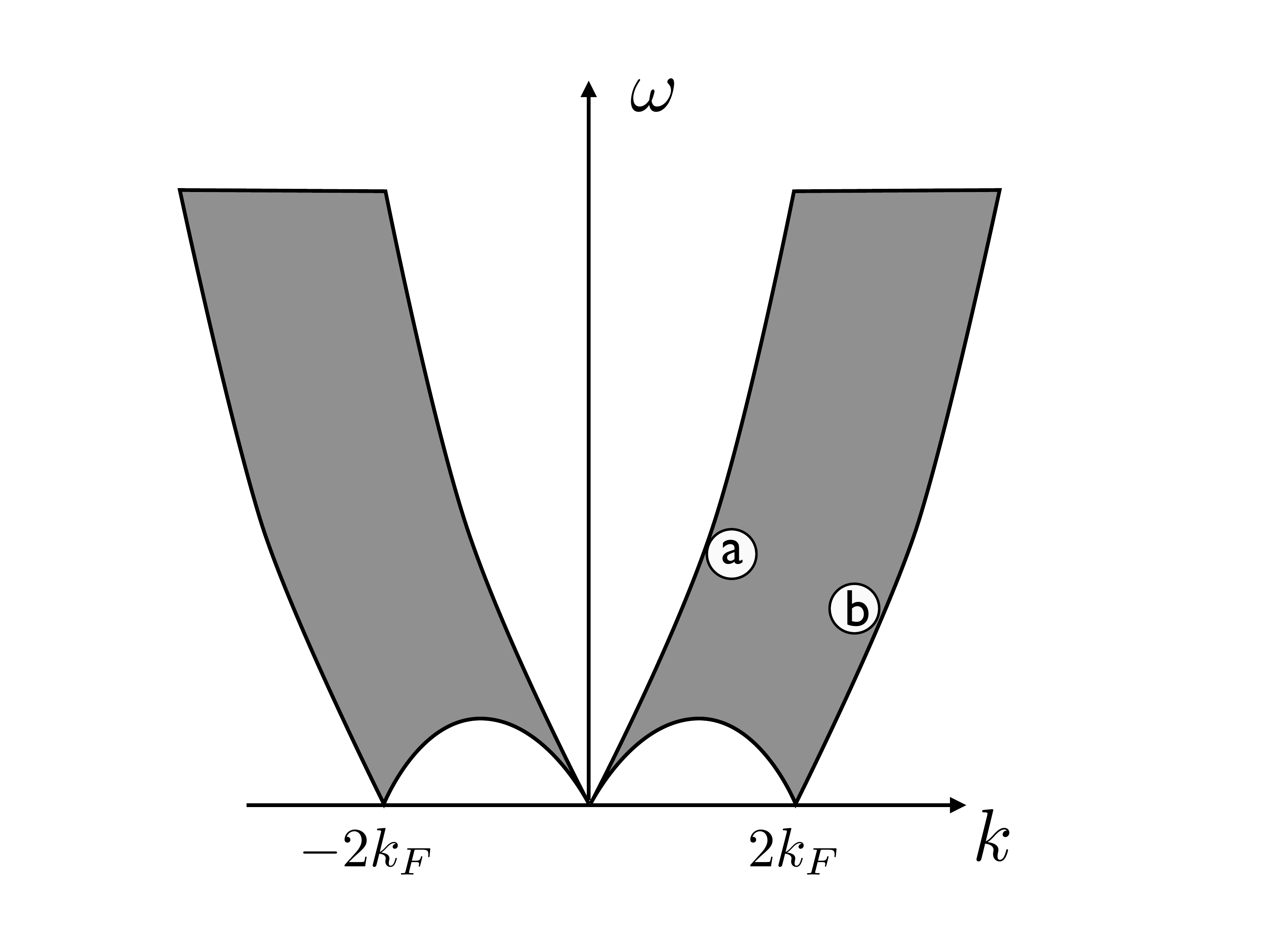}
\caption{Left: ground state for a generic system of noninteracting fermions in one dimension. The function $\varepsilon (k)$ is
the one-particle dispersion relation. The chemical potential $\mu$ sets the value of the Fermi wavevector $k_F$.
Filled single-particle states are represented by black dots, unfilled ones by open circles.
Right: one particle-hole excitation continuum. The a,b labels refer to the location within the continuum
of the particular single particle-hole excitations sketched in Fig. \ref{fig:1d_1ph_exc}.
}
\label{fig:1d_Fermi_sea}
\end{figure}
\end{center}

\begin{center}
\begin{figure}
\includegraphics[width=8cm]{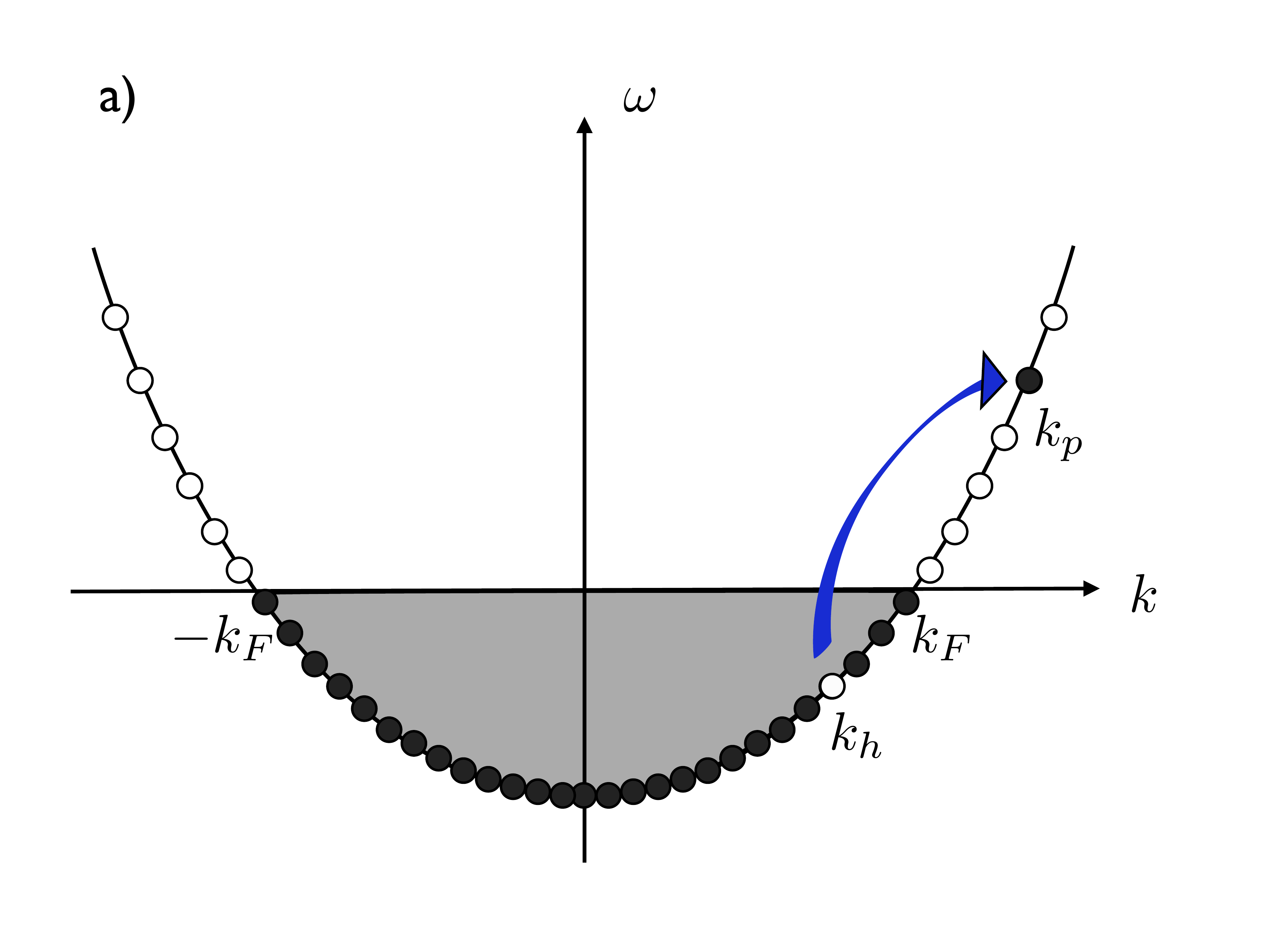}
\includegraphics[width=8cm]{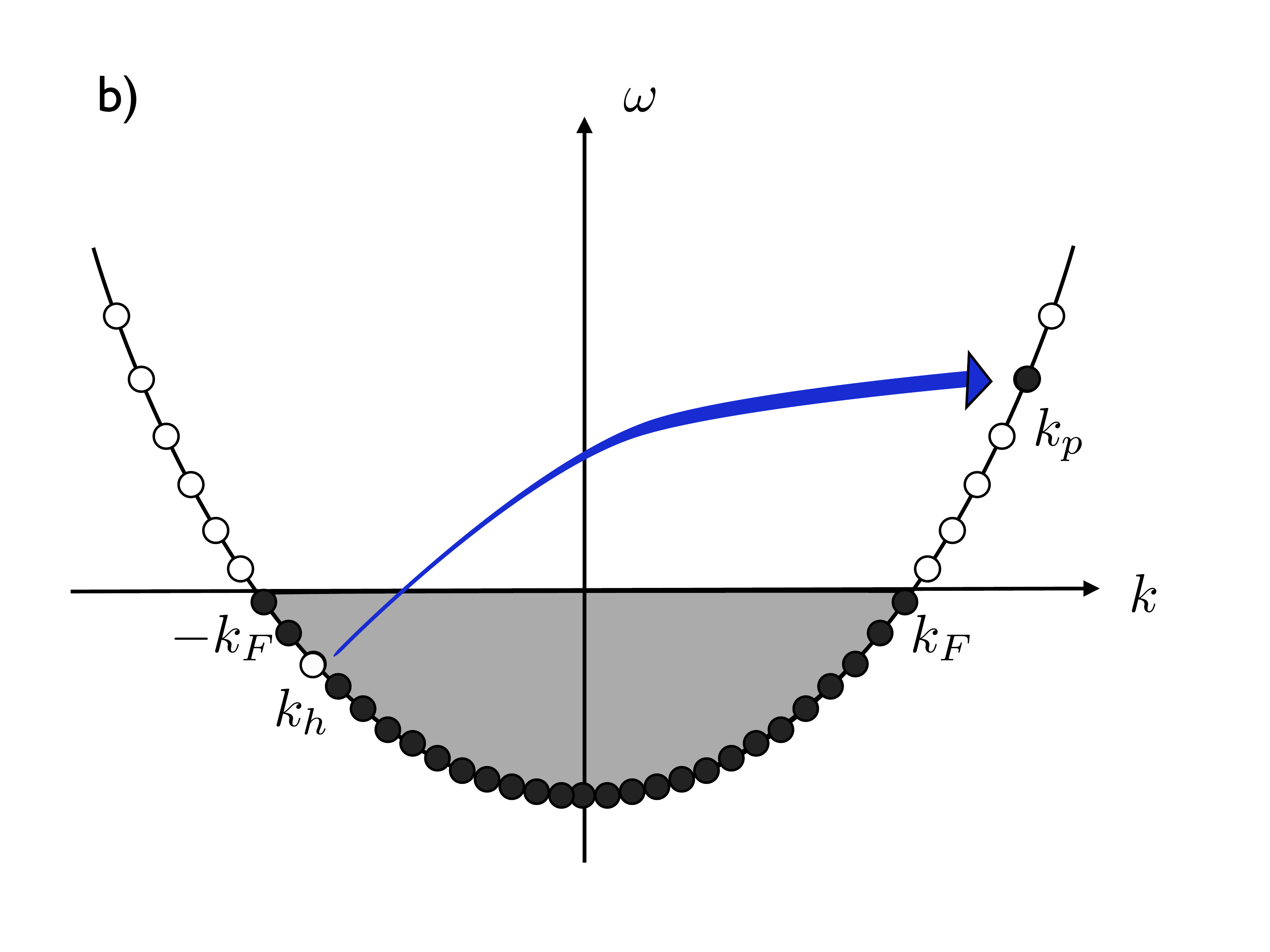}
\caption{Simple examples of one particle-hole excitations. The a,b labels refer to the right panel of Fig. \ref{fig:1d_Fermi_sea}.}
\label{fig:1d_1ph_exc}
\end{figure}
\end{center}

\begin{center}
\begin{figure}
\includegraphics[width=14cm]{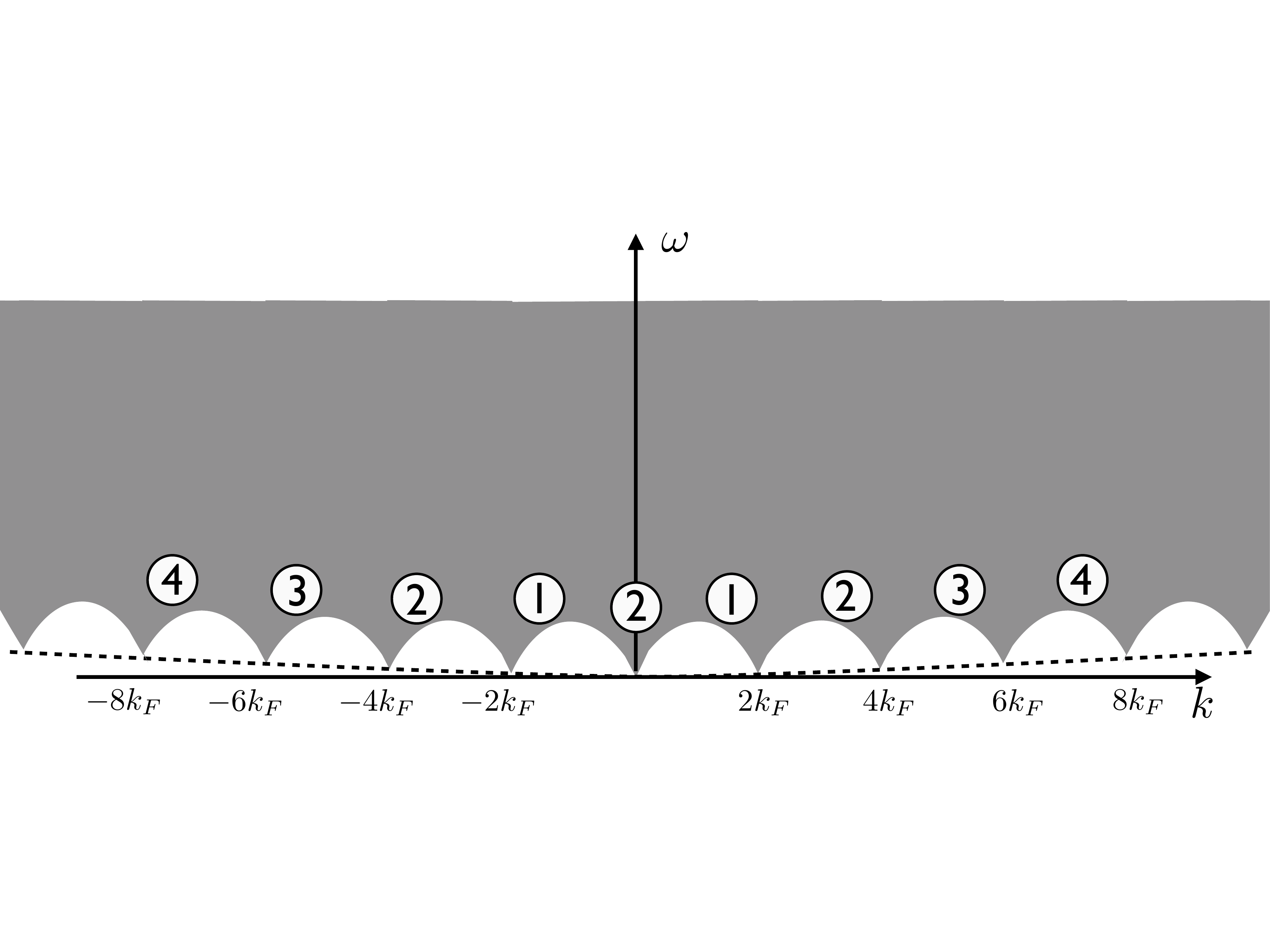}
\caption{Multiple particle-hole continuum. The numerals indicate the minimal number of particle-hole
excitations needed for a low-energy state to be found in this vicinity. The dashed line indicates
the minimal energy parabola for multiple Umklapp states on an exaggerated scale (this is of order $1/L$ for small
numbers of Umklapps and is typically neglected).
}
\label{fig:1d_multiph_continuum}
\end{figure}
\end{center}

At the simplest level, one-dimensionality alters the kinematics of simple excitations around a Fermi sea (Fig. \ref{fig:1d_Fermi_sea}).
The spectrum is formed by particle/hole excitations in the vicinity of the two Fermi points (Fig. \ref{fig:1d_1ph_exc}), and their
(multiple) Umklapp modes. In 1d, in contrast to higher dimensions, forbidden
regions of the frequency $\omega$-momentum $k$ plane exist. Low-energy modes around integer multiples of the Fermi momentum $k_F$ are always present (Fig. \ref{fig:1d_multiph_continuum}), but there
exist no low-energy excitations in the lobes between points $2(j-1) k_F$ and $2j k_F$ for any integer $j$.
For energies (much) below the height of the lobes (so $\omega \ll v_F k_F$ in which $v_F \equiv \frac{d \varepsilon(k)}{dk}|_{k_F}$
is the Fermi velocity), we can thus explicitly separate the excitations into different sectors labelled by an even integer $J = 2j$.

The lowest-energy state of sector $J=2j$ is the $j$-th Umklapp state at momentum $2j k_F$. These states are `persistent current' modes obtained by a Galilean transformation of
the ground state, giving momentum $2jk_F$ to the whole system, and carrying a quadratic (in $j$) energy shift.

Above the persistent current Umklapp states, but for energies still low on the scale of $v_F k_F$,
the spectrum of the theory is expected to be given by a linear in momentum boson (sound wave) spectrum (up to
nonlinear corrections).
Similarly, adding $\tilde{N}$ particles
shifts the Fermi momentum linearly (in $\tilde{N}$), and the energy quadratically.
We can thus immediately guess, following Haldane's reasoning in \cite{1981_Haldane_JPC_14},
that an effective theory for our fermions universally takes the form
\begin{equation}
H = E_{GS} + \sum_q v_S |q| b^\dagger_q b_q + \frac{2\pi}{L} (v_N \tilde{n}^2 + v_J j^2)
\label{eq:Haldaneguess}
\end{equation}
where $E_{GS}$ is the ground-state energy and $v_S, v_N$ and $v_J$ respectively denote
sound, charge and current velocities. It was Haldane's great insight to realize that in the presence of
interactions, the effective description based on Eq. (\ref{eq:Haldaneguess}) remains valid,
albeit with renormalized
parameters $v_S, v_N$ and $v_J$ (constrained by $v_s^2 = v_N v_J$).
This theory, which is thus universally valid for gapless one-dimensional systems (be they based on underlying
fermionic, bosonic or spin degrees of freedom), should be viewed as a `theory of everything' in 1d, similarly
to the status that Landau's Fermi liquid theory has achieved in higher dimensions. This is the point of view
so eloquently put forward in Haldane's paper \cite{1981_Haldane_JPC_14}.

Many important contributions paved the way for this synthesis.
The development of bosonization goes back almost to the very beginnings of quantum mechanics.
In 1934, Bloch \cite{1934_Bloch_HPA_7} used the fact that 1d fermions have the same type of low-energy excitations
as a harmonic chain in his study of incoherent X-ray diffraction.
Some years later, in 1950, Tomonaga \cite{1950_Tomonaga_PTP_5} applied Bloch's sound wave method to interacting fermions in 1d.
His main contribution was probably to realize that the physical density operator splits up into left-
and right-moving modes, obeying a bosonic Hamiltonian. He however missed the $2k_F$ contribution to the density correlation function,
and did not notice the anomalous decay of correlations.
The fundamental paper of Luttinger \cite{1963_Luttinger_JMP_4} follows in 1963, in which (perhaps unaware of Tomonaga's 1950 paper)
he formulated his (part)namesake model. Using Toeplitz determinants, he found that the average occupation in the
ground state behaves as a power-law with an
interaction-dependent anomalous dimension. He thus realized the crucial fact that the Fermi surface discontinuity
is destroyed by interactions in 1d.
This work was however incorrect in its treatment of the commutation relations of the density operators.
Later, Mattis and Lieb offered a correct treatment in their seminal paper \cite{1965_Mattis_JMP_6}.

The idea of bosonization, namely that bosons could be used to construct a complete set of states of
a 1d fermionic system, appeared in 1965 in \cite{1965_Overhauser_P_1}.
Subsequently, early computations of correlation functions appeared in \cite{1967_Theumann_JMP_8} and \cite{1968_Dover_AP_50}.
In particular, in the first of these, Theumann noticed the absence of single-particle poles in the Green's function.
She thus correctly concluded that single-particle excitations are absent in such theories.
In their famous 1974 paper \cite{1974_Dzyaloshinskii_JETP_38}, Dzyaloshinskii and Larkin recovered the
absence of single-particle pole and of Fermi surface discontinuity, and offered an
interpretation of Mattis and Lieb's solution starting from conventional diagrammatic perturbation theory.

An early version of the actual bosonization operator identity appeared in \cite{1969_Schotte_PR_182}.
This was refined by Mattis in 1974 \cite{1974_Mattis_JMP_15}, rendering calculation of correlation functions straightforward.
Similar results appeared in the work of Luther and Peschel \cite{1974_Luther_PRB_9}.
The power-law form for correlations was also recovered from equations of motion techniques in \cite{1974_Everts_SSC_15}.
Bosonization was then applied to spin chains and vertex models in \cite{1975_Luther_PRB_12}.
The first precise field-theoretical bosonization formula (as an operator identity) including the (until that point neglected)
particle number raising/lowering operators (Klein factors) is in general attributed to \cite{1975_Heidenreich_PLA_54}.
An early formulation of Luttinger liquid concepts appeared in 1975 in \cite{1975_Efetov_JETP_42}.

But the fact remains that it is Haldane, in a remarkable series of papers,
who gave Luttinger liquid theory the form it has today.
Starting in \cite{1979_Haldane_JPC_12}, he gave the first explicit construction of charge-raising operatos (Klein factors).
Subsequently, in \cite{1980_Haldane_PRL_45}, \cite{1981_Haldane_PRL_47} and most notably in \cite{1981_Haldane_JPC_14} he offered the
complete and explicit construction of the bosonization operator identities, and cross-checked results
with exactly solvable models. Most importantly, he proposed the concept of the Luttinger liquid (as he so defined it) as the
proper replacement for the Fermi liquid in one dimension, and showed that many different types of systems of fermions, bosons and spins
belong to this new universality class. This realization sparked much of the revolutionary advances achieved in low-dimensional quantum systems over the last 35 years.

\vspace{8mm}

\bibliographystyle{unsrt}


\begin{thebibliography}{10}

\bibitem{1981_Haldane_JPC_14}
F.~D.~M. Haldane.
\newblock {`Luttinger liquid theory' of one-dimensional quantum fluids. I.
  Properties of the Luttinger model and their extension to the general 1D
  interacting spinless Fermi gas}.
\newblock {\em J. Phys C: Sol. St. Phys.}, 14(19):2585, 1981.

\bibitem{1957_Landau_JETP_3}
L.~D. Landau.
\newblock {Theory of Fermi-liquids}.
\newblock {\em Sov. Phys. JETP}, 3:920, 1957.

\bibitem{1957_Landau_JETP_5}
L.~D. Landau.
\newblock {Oscillations in a Fermi-liquid}.
\newblock {\em Sov. Phys. JETP}, 5:101, 1957.

\bibitem{AbrikosovBOOK}
A.~A. Abrikosov, L.~P. Gorkov, and I.~E. Dzyaloshinski.
\newblock {\em {Methods of Quantum Field Theory in Statistical Physics}}.
\newblock Dover, 1963.

\bibitem{1934_Bloch_HPA_7}
F.~Bloch.
\newblock {Inkoh\"arente R\"ontgenstreuung und Dichteschwankungen eines
  entarteten Fermigases}.
\newblock {\em Helv. Phys. Acta}, 7:385, 1934.

\bibitem{1950_Tomonaga_PTP_5}
S.-I. Tomonaga.
\newblock {Remarks on Bloch's Method of Sound Waves applied to Many-Fermion
  Problems}.
\newblock {\em Prog. Theor. Phys.}, 5(4):544--569, 1950.

\bibitem{1963_Luttinger_JMP_4}
J.~M. Luttinger.
\newblock {An Exactly Soluble Model of a Many-Fermion System}.
\newblock {\em Journal of Mathematical Physics}, 4(9):1154--1162, 1963.

\bibitem{1965_Mattis_JMP_6}
D.~C. Mattis and E.~H. Lieb.
\newblock {Exact Solution of a Many-Fermion System and Its Associated Boson
  Field}.
\newblock {\em J. Math. Phys.}, 6(2):304--312, 1965.

\bibitem{1965_Overhauser_P_1}
A.~W. Overhauser.
\newblock Note on the band theory of magnetism.
\newblock {\em Physics}, 1:307, 1965.

\bibitem{1967_Theumann_JMP_8}
A.~Theumann.
\newblock {Single-Particle Green's Function for a One-Dimensional Many-Fermion
  System}.
\newblock {\em J. Math. Phys.}, 8(12):2460--2467, 1967.

\bibitem{1968_Dover_AP_50}
C.~B. Dover.
\newblock {Properties of the Luttinger model}.
\newblock {\em Ann. Phys. (NY)}, 50(3):500 -- 533, 1968.

\bibitem{1974_Dzyaloshinskii_JETP_38}
I.~E. Dzyaloshinskii and A.~I Larkin.
\newblock {Correlation functions for a one-dimensional Fermi system with
  long-range interaction (Tomonaga model)}.
\newblock {\em Sov. Phys. JETP}, 38, 1974.
\newblock Russian original: ZhETF, Vol. 65, No. 1, p. 411, January 1974.

\bibitem{1969_Schotte_PR_182}
K.~D. Schotte and U.~Schotte.
\newblock {Tomonaga's Model and the Threshold Singularity of X-Ray Spectra of
  Metals}.
\newblock {\em Phys. Rev.}, 182(2):479--482, 1969.

\bibitem{1974_Mattis_JMP_15}
D.~C. Mattis.
\newblock {New wave-operator identity applied to the study of persistent
  currents in 1D}.
\newblock {\em J. Math. Phys.}, 15(5):609--612, 1974.

\bibitem{1974_Luther_PRB_9}
A.~Luther and I.~Peschel.
\newblock {Single-particle states, Kohn anomaly, and pairing fluctuations in
  one dimension}.
\newblock {\em Phys. Rev. B}, 9(7):2911--2919, 1974.

\bibitem{1974_Everts_SSC_15}
H.~U. Everts and H.~Schulz.
\newblock {Application of conventional equation of motion methods to the
  Tomonaga model}.
\newblock {\em Solid State Commun.}, 15(8):1413 -- 1416, 1974.

\bibitem{1975_Luther_PRB_12}
A.~Luther and I.~Peschel.
\newblock Calculation of critical exponents in two dimensions from quantum
  field theory in one dimension.
\newblock {\em Phys. Rev. B}, 12(9):3908--3917, 1975.

\bibitem{1975_Heidenreich_PLA_54}
R.~Heidenreich, B.~Schroer, R.~Seiler, and D.~Uhlenbrock.
\newblock {The Sine-Gordon equation and the one-dimensional electron gas}.
\newblock {\em Phys. Lett. A}, 54(2):119 -- 122, 1975.

\bibitem{1975_Efetov_JETP_42}
K.~B. Efetov and A.~I. Larkin.
\newblock {Correlation functions in one-dimensional systems with a strong
  interaction}.
\newblock {\em Sov. Phys. JETP}, 42(2):390, 1975.

\bibitem{1979_Haldane_JPC_12}
F.~D.~M. Haldane.
\newblock {Coupling between charge and spin degrees of freedom in the
  one-dimensional Fermi gas with backscattering}.
\newblock {\em J. Phys. C: Solid State Phys.}, 12(22):4791, 1979.

\bibitem{1980_Haldane_PRL_45}
F.~D.~M. Haldane.
\newblock {General Relation of Correlation Exponents and Spectral Properties of
  One-Dimensional Fermi Systems: Application to the Anisotropic $S=\frac{1}{2}$
  Heisenberg Chain}.
\newblock {\em Phys. Rev. Lett.}, 45(16):1358--1362, 1980.

\bibitem{1981_Haldane_PRL_47}
F.~D.~M. Haldane.
\newblock {Effective Harmonic-Fluid Approach to Low-Energy Properties of
  One-Dimensional Quantum Fluids}.
\newblock {\em Phys. Rev. Lett.}, 47(25):1840--1843, 1981.

\end{thebibliography}

\end{document}